\documentclass[runningheads]{llncs}
\usepackage[T1]{fontenc}
\usepackage{graphicx}
\usepackage{xcolor}
\usepackage{url}
\usepackage{hyperref}
\hypersetup{urlcolor=blue}
\usepackage{color}

\usepackage{comment}
\usepackage[nohyperlinks,printonlyused,withpage]{acronym}
\usepackage{cleveref}
\usepackage{makecell}
\usepackage[table]{xcolor}

\begin{document}
\title{(EC)2: Event-Centric Explainability for Cybersecurity Through Multi-Agent LLM Investigations}
\titlerunning{Event-Centric Explainability for Cybersecurity}
% If the paper title is too long for the running head, you can set
% an abbreviated paper title here
%
% \author{Neta Kirmayer \and
% Rami Puzis \and
% Asaf Shabtai \and
% David Tayouri}

\author{Neta Kirmayer\inst{1} \and
David Tayouri\inst{1} \and
Andrés Murillo\inst{2} \and
Motoyoshi Sekiya\inst{2} \and
Asaf Shabtai\inst{1} \and
Rami Puzis\inst{1}
}

% Third Author\inst{3}\orcidID{2222--3333-4444-5555}}
% %
\authorrunning{N. Kirmayer et al.}
% % First names are abbreviated in the running head.
% % If there are more than two authors, 'et al.' is used.
% %
\institute{Ben-Gurion University of the Negev, Beer-Sheva, Israel \and
Fujitsu}
% Springer Heidelberg, Tiergartenstr. 17, 69121 Heidelberg, Germany
% \email{lncs@springer.com}\\
% \url{http://www.springer.com/gp/computer-science/lncs} \and
% ABC Institute, Rupert-Karls-University Heidelberg, Heidelberg, Germany\\
% \email{\{abc,lncs\}@uni-heidelberg.de}}
%
\maketitle              % typeset the header of the contribution
\begin{abstract}
Security operations centers rely on anomaly detection systems to flag suspicious events. 
Feature-level explanations for anomaly detectors offer limited value for operational investigations. 
To effectively handle alerts, analysts need to know contextual relationships and need actionable understanding of the entities involved.
This paper introduces an event-centric detector-agnostic approach for explaining cybersecurity alerts in small- to medium-sized enterprise networks.
We present (EC)2, a multi-agent framework that performs structured, hypothesis-driven investigation to provide explanations grounded in verifiable evidence.
Evaluation results show that the proposed framework improves post-detection analysis by generating operationally meaningful explanations, which also enhance event classification accuracy.

\end{abstract}

\section{Introduction}

As cybercrime continues to flourish, organizations are under increasing pressure to quickly and effectively respond to security incidents~\cite{cichonski2012computer}. 
\Acp{SOC} help organizations monitor and manage cybersecurity threats and incidents. 
To monitor and respond to threats, \acp{SOC} employ tools and dashboards to capture and identify security-relevant events in real time, presenting them to internal security analysts for severity and relevance assessment~\cite{vielberth2020security}.
As cyberattacks have increased in volume and sophistication, the workload of \ac{SOC} analysts has grown due to the rising number of alerts, volume of data to be reviewed, and scope of investigation required~\cite{tariq2025alert}.

While automation and advanced analytics have helped \acp{SOC} manage the growing number of incidents, they introduce a new issue: many detectors now incorporate \ac{ML} models whose internal reasoning is opaque. 
Despite their effectiveness in identifying threats, \ac{ML}-based systems provide limited insight into the relationship between flagged events and the broader attack context \cite{zhang2022explainable}. 
This interpretability gap is particularly problematic in forensic investigation, which relies on understanding causal relationships, reconstructing event sequences, and tracing the roles of network entities, requirements that go far beyond mere detection accuracy.

The field of \ac{XAI} aims to address this gap. 
Traditional \ac{XAI} methods provide feature-level explanations that identify which input attributes influenced a model's decision.
However, when applied to the \ac{ML}-based detectors employed in \acp{SOC}, these methods have notable limitations.
While feature attribution clarifies \textit{what influenced a model's decision}, it does not explain \textit{why an event is suspicious}.
The knowledge that a model flagged a network flow due to a high packet rate or uncommon port usage does not indicate whether the event is part of an attack or a benign anomaly.
When investigating security incidents, analysts must understand causal relationships, temporal sequences, and cross-event dependencies - contextualized insights that feature-level explanations alone cannot provide.
%First, automated detectors frequently produce false positives, flagging benign events as suspicious; this increases the investigative workload, as analysts must manually validate each alert \cite{islam2023application}.
%This gap forces analysts to perform time-consuming manual investigations to understand the reasoning behind alerts, increasing response time and undermining trust in automated detection systems.

To address this challenge, we adopt a novel event-centric approach to explainability.
Our approach specifically targets the interpretability gap by providing visibility into the investigative logic that connects anomalous observations to potential causes.
We address the post-detection phase of the \ac{SOC} workflow, in which analysts must understand, validate, and contextualize detected anomalies yet often lack clear explanations of the reasoning behind them.
%Rather than explaining a model's decision, our framework explains the security event itself through hypothesis-driven analysis.
We present (EC)2, a multi-agent framework in which \ac{LLM}-based agents collaboratively conduct hypothesis-driven investigations: when a detector raises an alert, (EC)2 takes the triggering event as input and produces a structured report that summarizes the hypotheses considered, the evidence gathered, and the conclusions reached.
By explaining why an event is anomalous rather than why a model flagged it, (EC)2 provides operationally meaningful support for decision-making.
%By organizing the reasoning process into distinct investigative roles, (EC)2 produces structured, interpretable explanations that mirror an analyst’s thought process, helping reduce their cognitive load and increasing confidence in automated detections. 

(EC)2 fits into XAI as a post-hoc, model-agnostic, local explanation method.
It is post-hoc in that it analyzes events after a separate detector has flagged them, without relying on that model's internals.
It is model-agnostic, because it can investigate anomalies produced by any type of detector - rule-based, statistical, or deep learning. %, as long as the detector designates the event as suspicious.
It is a local explanation method: each run of (EC)2 examines a single anomalous event and reconstructs the specific context and behaviors that render it unusual.

The proposed framework uses \ac{RAG} to ensure that every explanation and investigative hypothesis is grounded in real contextual evidence.
This links reasoning steps to heterogeneous verifiable data sources, reducing uncertainty and increasing analysts' trust in the explanations generated.
The framework's multi-agent paradigm provides a structural foundation for implementing investigative reasoning in cybersecurity.
Rather than relying on a single central model, (EC)2 distributes reasoning across multiple interacting agents that collaborate to analyze evidence, form conclusions, and refine understanding.
By externalizing reasoning as a transparent, interactive dialogue, the multi-agent architecture enhances interpretability and enables analysts to understand how conclusions are reached through explicit, evidence-based explanations.
To demonstrate and evaluate our framework, we apply it to network anomaly analysis, using datasets containing both malicious and benign events.
We show that event-centric explanations provide substantially greater interpretive and operational value than feature-based attribution explanations.
In summary, we make the following contributions:
\begin{enumerate}
    \item We introduce an event-centric XAI approach for cybersecurity that explains why the event itself is anomalous rather than why a detector made a particular decision.
    \item We develop a fully automated, hypothesis-driven multi-agent investigation framework that generates hypotheses, retrieves evidence from heterogeneous data sources, and produces structured investigation reports grounded in verifiable evidence.
    \item We provide an empirical comparison between (EC)2 and a state-of-the-art LLM-based explainability system for network intrusion detection, evaluating both approaches on the same set of anomalous events and across diverse attack scenarios. % using a structured, multi-criteria explanation quality framework.
    Our results demonstrate that (EC)2 produces substantially richer and more operationally meaningful explanations across all evaluated attack categories and quality criteria.
\end{enumerate}

\begin{comment}
The structure of this paper is as follows: \Cref{sec:background} introduces the technical background and identifies the gaps in existing methods that motivate this work.
In \Cref{sec:system architecture}, we present the proposed system architecture.
The experimental setup and the dataset used in the evaluation are outlined in \Cref{sec:experimentalSetup}.
\Cref{sec:evaluation} details the evaluation methodology and discusses the results. \Cref{sec:relatedWork} reviews prior research and places our contribution in relation to existing approaches.
Finally, \Cref{sec:conclusions} summarizes the main insights of the study and highlights promising directions for future work.
\end{comment}

\section{Background} \label{sec:background}
\subsection{Network Anomaly Detection}
Anomaly detection identifies patterns in data that deviate from expected behavior.
In cybersecurity, this capability is essential for discovering evolving attacks that may not match existing signatures \cite{ahmed2025signature}.
In network contexts, anomaly detectors learn a baseline of normal activity and flag deviations as potential threats, typically operating on feature representations derived from raw traffic data such as packet captures or flow records.
These representations may capture attributes such as packet-size distributions, protocol usage, connection durations, or communication frequency between hosts, and are used to train \ac{ML} or \ac{DL} under supervised, unsupervised, or hybrid learning paradigms.
%To capture the complexity of real-world network behavior, increasingly sophisticated models have been proposed, ranging from ensemble techniques and autoencoders to graph-based architectures capable of learning nonlinear and relational patterns \cite{bhuyan2013network,mirsky2018kitsune}.
%While such models improve detection accuracy, their complexity often reduces interpretability, making it difficult to understand why specific events were classified as anomalous.

\textbf{Anomaly Detection in SOCs}.
A \acf{SOC} is the organizational unit responsible for continuously monitoring, detecting, and responding to cybersecurity incidents across an enterprise \cite{chamkar2024security}.
\ac{SOC} teams employ human expertise, defined workflows, and automated tools to manage alerts and coordinate responses.
Integrating anomaly detection into \ac{SOC} operations introduces two major issues.
First, because anomaly detectors flag deviations from normal behavior, they often misidentify benign activities as anomalies, leading to high false-positive rates and increased investigative burden \cite{ali2024unveiling}.
Second, even when a detected anomaly corresponds to a real attack, many models offer little transparency into why a particular event was flagged, reducing analyst trust, complicating investigations, and delaying response actions \cite{sarker2024explainable}.

\subsection{Explainable Artificial Intelligence (XAI)}
The growing complexity of modern \ac{ML} and \ac{DL} models has exacerbated the interpretability problem, as they often offer little transparency into their internal decision-making.
This issue is particularly important in cybersecurity, where analysts must understand a model's reasoning to trust its output and act effectively on it, and where false alarms or missed detections can have severe consequences~\cite{zhang2022explainable,arrieta2020explainable}.
\Ac{XAI} address this by developing methods that make ML models more interpretable and understandable to humans \cite{arrieta2020explainable}.

\ac{XAI} methods can be categorized along several key dimensions.
\textbf{Ante-hoc} approaches involve models that are interpretable by design, such as decision trees, logistic regression, and rule-based models \cite{retzlaff2024post}.
\textbf{Post-hoc} methods explain black-box models after training; prominent examples include LIME \cite{ribeiro2016should} and SHAP \cite{lundberg2017unified}, which output ranked features with contribution scores.
\textbf{Model-specific} methods exploit a model's internal structure to generate explanations, such as gradients in neural networks or feature splits in decision trees, while \textbf{model-agnostic} methods treat the model as a black box and rely solely on its inputs and outputs, enabling broad applicability across different model types \cite{arrieta2020explainable}.
\textbf{Local} methods explain individual predictions by identifying which features most influenced a specific output, while \textbf{global} methods describe a model's overall behavior across an entire dataset \cite{das2020opportunities}.

\subsection{LLMs for Cybersecurity and Explainability}
The emergence of \acp{LLM} such as GPT \cite{radford2018improving} and Gemini \cite{team2023gemini} has transformed \ac{AI}, enabling sophisticated language understanding and reasoning capabilities.
In cybersecurity operations, \acp{LLM} support human analysts by processing vast amounts of textual data, enriching alerts with contextual information, and reducing cognitive workload by performing tasks such as log summarization, automated alert triage, and vulnerability analysis \cite{bandyopadhyay2025thinking,srinivas2025ai}.
Traditional XAI methods often fail to provide the causal and contextual reasoning needed for effective \ac{SOC} decision-making.

\section{(EC)2 Architecture} \label{sec:system architecture}

(EC)2's primary objective is to provide human-understandable explanations for network events flagged as anomalous.
% As discussed above, 
Traditional feature-based XAI systems are insufficient in the cybersecurity domain. While they can highlight which features are important, they fail to explain why an event is suspicious, forcing analysts to conduct time-consuming manual investigations.
To address this limitation, our framework produces natural language explanations, the format most accessible and actionable for human responders. As input, (EC)2 receives a single anomalous event, typically a flagged packet or NetFlow generated by a base anomaly detector, and produces a structured report that documents the investigation process and the evidence-based reasoning through which the anomaly is explained.
% (EC)2 performs a structured investigation that reflects how analysts review evidence, relate behaviors, and reach conclusions.

At the core of this approach lies a hypothesis-driven investigation process.
(EC)2 formulates plausible hypotheses for why an event may be anomalous and iteratively gathers evidence to confirm or refute each, continuing until a coherent and verifiable conclusion is reached.
The investigation unfolds as a sequence of analytical steps: hypothesis formulation, evidence retrieval, cross-validation, and explanation generation that closely resembles the reasoning pattern used by human analysts.
In practice, (EC)2 reproduces the actions a human analyst would take during an investigation.
% An analyst would typically consult packet captures, log repositories, or network topology data to verify a suspicion.
% (EC)2 mirrors this by automatically querying the same data sources.
Each intermediate decision is grounded entirely in the evidence extracted from external data sources, ensuring that the final explanation is both logically consistent and empirically supported.
%These data sources, therefore, play a foundational role in the investigation - the relevance and quality of the evidence (EC)2 can retrieve depends directly on what is made available to it. 
% In network anomaly analysis, two natural sources of such context exist: network topology and configuration data, which captures the structure and known behavior of the enterprise network, and network traffic data, which records observed activity across it. Together, these sources provide the contextual grounding necessary for producing reliable and accurate explanations.
Leveraging these data sources, (EC)2 performs the investigation through a coordinated multi-agent workflow driven by an \ac{LLM}.
Because the investigation requires multiple levels of reasoning, (EC)2 distributes these responsibilities across several specialized agents, each interacting with the \ac{LLM} independently to perform a specific analytical role. %, rather than consolidating all reasoning into a single context.
By orchestrating these agents into a structured workflow, we ensure that the investigation is well-organized, transparent, and consistently grounded in the verifiable evidence retrieved from the data sources.
The use of LLM-based agents enables flexible, context-sensitive reasoning while strictly complying with (EC)2's evidence-driven methodology.

\subsection{Data and Evidence Sources} \label{subsec:approach:data_and_evidence}
The investigation framework relies on multiple heterogeneous data sources, each providing a distinct perspective on the network environment.
% To ensure that explanations are grounded in factual evidence rather than model-generated assumptions, (EC)2 adopts a \ac{RAG} mechanism, in which the LLM agents do not rely only on their internal knowledge.
% Instead, they retrieve information from external data stores and base their reasoning on the results returned.
To ensure that explanations are grounded in factual evidence rather than model-generated assumptions, (EC)2 adopts a RAG mechanism: agents do not rely on internal knowledge but retrieve information from external data sources and base their reasoning on the results returned.
Employing \ac{RAG} in the investigation pipeline ensures that each analytical step is supported by verifiable evidence.
% To maintain a clear division of responsibilities in the multi-agent architecture, each data source is assigned to a single agent.
% We refer to any agent with direct access to an external data source as a retrieval agent.
% This design guarantees that only one agent can retrieve information from a given source, preventing overlap or inconsistencies.
% Other agents rely on the retrieval agent's output.
% Because (EC)2 interacts with different types of information, each retrieval agent is equipped with a dedicated tool designed for its assigned data source. 
% These tools are functions that translate agent queries into appropriate database operations and return structured results.
% This architecture also plays a critical role in controlling hallucination. Since a retrieval agent can only reason about evidence that it explicitly retrieves through its tool, it cannot fabricate data to support a hypothesis or align with the ongoing investigation.
To maintain a clear division of responsibilities, each data source is assigned to a dedicated retrieval agent equipped with a callable tool - a function the agent invokes during reasoning to translate queries into appropriate database operations and return structured results. Other agents rely solely on the retrieval agent's output. This design also plays a critical role in controlling hallucination: since a retrieval agent can only reason about evidence it explicitly retrieves through its tool, it cannot fabricate data to support a hypothesis or align with the ongoing investigation.
As a result, the framework enforces an evidence-driven workflow in which all reasoning must be anchored to evidence retrieved from the underlying data sources.

In practice, each retrieval agent operates behind a natural language interface.
The agents receive questions phrased in natural language and return natural-language answers, without exposing the underlying queries or raw database results to the other agents.
% All specialized operations, parsing of structured outputs, and their conversion into readable natural language take place entirely within the agent's internal process.
% This design keeps the multi-agent conversation human-interpretable, while ensuring that every response is grounded in the evidence retrieved from the corresponding data source.
All structured operations and output parsing take place entirely within the agent's internal process, keeping the multi-agent conversation human-interpretable while ensuring every response is grounded in retrieved evidence.

For our network anomaly explanation, we rely on two different data sources:

\textbf{Network Topology:} To represent the structure of the enterprise network, the framework uses a knowledge graph - a graph-based topology model.
In this graph, nodes correspond to devices and capture their configuration details, while edges encode the connections and relationships between them.
This design supports reasoning about the relational and structural aspects of the environment, including how devices are connected, how traffic can flow, and which entities are relevant to a given anomalous event.
In our framework, the network topology is stored in a Neo4j graph database. The dedicated retrieval agent communicates with this data source by generating Cypher queries (Neo4j's query language).
The agent receives natural language questions such as: "What is the MAC address of the device with IP address X.X.X.X?"
Internally, it constructs the appropriate Cypher query, issues it to the graph database, retrieves the results, processes the structured output, and returns a concise natural language answer based solely on the information stored in the knowledge graph.

\textbf{Network Traffic:} To support investigation of network behavior, raw traffic data is transformed into structured tabular representations stored in a relational database. The specific representations constructed depend on the available data and dataset characteristics.
This tabular representation enables EC(2) to reason about temporal and behavioral patterns in traffic, such as packet sequences, communication flows, and deviations from normal activity.
The traffic database is stored in PostgreSQL, which provides efficient indexing and query execution over large packet sets.
The corresponding retrieval agent interacts with this data source through a natural language interface: it receives a question expressed in everyday language (e.g., "How many TCP packets were sent from IP address X.X.X.X to IP address Y.Y.Y.Y in the five packets before the event?").
The agent converts the question into a SQL query, executes it against the PostgreSQL database, and processes the resulting structured data.
The agent then returns a concise natural-language response derived solely from the retrieved evidence.

\vspace{-3mm}

\begin{figure*}[htp]
\centering
\includegraphics[width=\textwidth]{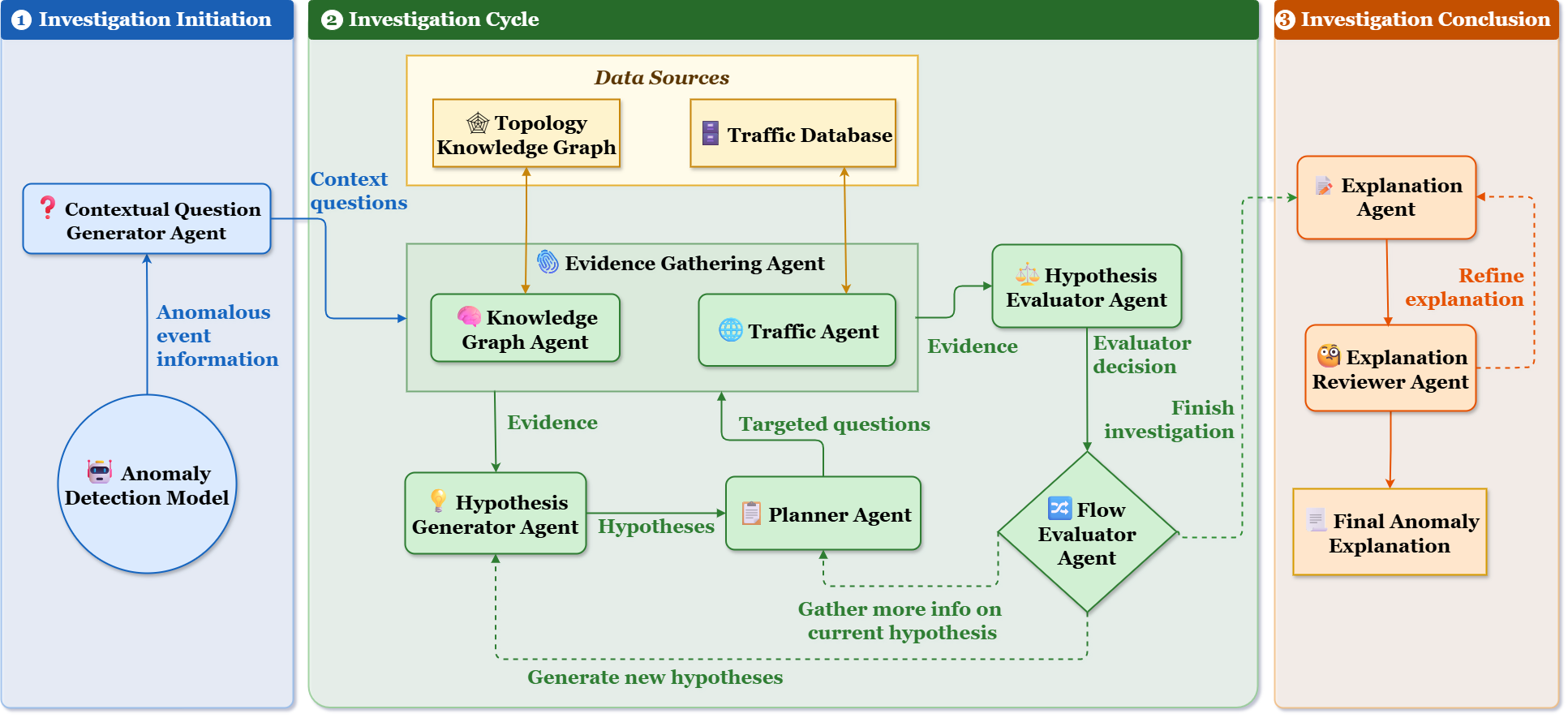} % full page width
\caption{(EC)2 architecture.}
\label{fig:full_workflow}
\end{figure*}

\vspace{-5mm}

% \vspace{-4mm}
\subsection{Agents and Workflow} \label{subsec:approach:agent_and_workflow}
(EC)2 is implemented using the AutoGen framework~\cite{autogenMSR}, in which each agent is configured with a distinct system prompt defining its role, reasoning scope, and output expectations.
Agents are organized into a directed graph that determines the execution order.
All agents share a common conversation context, meaning that each agent receives the full history of prior agents' interactions as part of its input, allowing it to build on previously gathered evidence and reasoning.
Most agents produce structured JSON outputs that serve two purposes: they help subsequent agents focus on the relevant parts of the conversation, and they enable AutoGen to parse agent outputs and apply conditional routing logic across graph transitions.
The \textit{Explanation Agent} and the \textit{Explanation Reviewer Agent} are exceptions that produce natural language narratives, as they are responsible for constructing the final human-readable report.

As illustrated in \Cref{fig:full_workflow}, the investigation process is divided into three main stages: Investigation Initiation, Investigation Cycle, and Investigation Conclusion. Each stage is described in detail below.

\vspace{-4mm}
\subsubsection{Investigation Initiation:} \label{subsubsec:approach:initiate_investigation}
When the anomaly detector identifies a network event as anomalous, the event information (raw packet metadata, NetFlow record, or any other representation of the flagged event) serves as the initial input to the multi-agent framework.
At this stage, no additional context is available. %: (EC)2 has only the single flagged event and must begin constructing an understanding of it from this minimal starting point.
The first component to operate on this input is the \textit{Contextual Question Generator Agent}.
Its role is to identify the factual information that is missing and must be retrieved before any hypothesis can be formed.
The agent receives the anomalous event information as input and produces a structured JSON output containing an ordered list of natural-language questions that target essential details to be gathered, such as device identity, address information, communication patterns, and specific event attributes.
The agent's sole purpose is to expand the factual context needed to begin an evidence-driven investigation.
All querying, schema structure, and technical constraints are handled behind the scenes, while the agent's output consists of a list of natural-language questions.
These questions are then passed to the \textit{Evidence Gathering Agent}, which resolves them by querying the appropriate data sources and returns a structured JSON list of question-answer pairs that forms the factual foundation for the next stage.

\vspace{-4mm}
\subsubsection{Investigation Cycle:} \label{subsubsec:approach:investigation_cycle}
After the relevant factual information has been retrieved, the framework enters the investigation cycle, which serves as the main analytical stage of (EC)2.
This stage includes generating, testing, and refining hypotheses that explain the anomalous event.
The cycle is iterative and continues until the investigation is concluded, either when the accumulated evidence sufficiently accounts for the anomalous event or when the maximum number of hypothesis generation rounds has been reached.

The cycle begins with the \textit{Hypothesis Generator Agent}, which proposes up to three candidate hypotheses to guide the investigation's next steps.
The agent receives the factual question-answer pairs compiled during the initiation stage and produces a structured JSON list of candidate hypotheses.
These hypotheses serve as investigative directions, highlighting potential explanations or avenues that require examination.
A hypothesis may describe a suspicious pattern, a possible misconfiguration, an unexpected device communication, a protocol inconsistency, or any other behavior that could meaningfully account for the anomalous event.
Importantly, the hypotheses are not limited to attack scenarios; they encompass all plausible investigative directions, including benign causes, environment-related anomalies, and configuration issues.

Each hypothesis proposed by the \textit{Hypothesis Generator Agent} is examined in turn.
The investigation for each hypothesis begins with the \textit{Planner Agent}, which receives the full conversation and produces a structured JSON output containing the current hypothesis under investigation, along with a list of targeted natural language questions whose answers will help assess the hypothesis.

These targeted questions are passed to the \textit{Evidence Gathering Agent}, which serves as the central coordinator of evidence retrieval throughout the cycle.
First employed in the initiation stage to resolve contextual questions, the \textit{Evidence Gathering Agent} continues to operate throughout the investigation cycle by routing the \textit{Planner Agent's} targeted questions to the appropriate retrieval agents and compiling their responses.
The \textit{Evidence Gathering Agent} receives the structured JSON question list along with the full conversation, and for each question, it determines which data source to query.
The agent then produces a structured JSON output comprising a list of question-answer pairs, each containing the original question along with the natural language answer provided by the corresponding retrieval agent.
For each question, those concerning device identity, relationships, and network structure are forwarded to the \textit{Knowledge Graph Agent}, whereas those about packet attributes, traffic behavior, and communication patterns are forwarded to the \textit{Traffic Agent}. The answers are then compiled into a structured list of question-answer pairs.
By centralizing all interactions with the data sources within the purview of the \textit{Evidence Gathering Agent}, the framework significantly reduces the burden on the other agents in the workflow - they do not need to know which data source is relevant to a given question or to perform query construction or schema reasoning.
This makes (EC)2 easily extensible: additional data sources can be integrated by adding new retrieval agents and updating only the \textit{Evidence Gathering Agent's} tools and prompts,  without requiring modifications to the other agents in the framework.

Each retrieval agent implements (EC)2's RAG mechanism, receiving a natural language question from the \textit{Evidence Gathering Agent} and answering it by executing the required queries against its respective database.
The gathered evidence is then reviewed by the \textit{Hypothesis Evaluator Agent}, which produces a structured JSON output containing the current hypothesis, its evaluation status (supported, refuted, or unresolved), and the reasoning behind that status.

Based on the \textit{Hypothesis Evaluator Agent's} structured output, the \textit{Flow Evaluator Agent} determines how the investigation should proceed by examining the entire conversation up to that point.
If the hypothesis is clearly supported or rejected, the \textit{Flow Evaluator Agent} moves the investigation on to the next hypothesis on the list. Multiple hypotheses may be supported within a single investigation, and all are included in the final explanation.
% If the evidence is deemed insufficient, the investigation returns to the \textit{Planner Agent} to request additional, more targeted information for the same hypothesis.
If the evidence is deemed insufficient, control returns to the \textit{Planner Agent}, which produces a revised set of targeted questions for the same hypothesis.
To prevent unbounded looping on a single hypothesis, this evidence-gathering loop is allowed to repeat at most once per hypothesis: after one additional cycle of planning, retrieval, and evaluation, the investigation for that hypothesis is considered complete, even if its final status remains only partially resolved.

This process is applied to every hypothesis generated.
After all hypotheses have been fully investigated, the \textit{Flow Evaluator Agent} examines the entire conversation to determine whether to terminate or continue the investigation.
At this point, the agent either concludes the investigation and shifts to the investigation conclusion stage if the existing evidence is sufficient, or triggers an additional round of hypothesis generation if the current hypotheses fail to explain the event.

% To prevent the investigation workflow from entering an unbounded loop of hypothesis generation, the framework allows returning to the \textit{Hypothesis Generator Agent} only once beyond the initial hypothesis generation. Consequently, (EC)2 can produce at most two sets of hypotheses: the initial set and, if necessary, one additional set.

This design allows the framework to explore alternative explanations when the correct attack hypothesis is not included in the initial set, while keeping the investigation process bounded and manageable. 
If a second round of hypotheses was generated, the \textit{Flow Evaluator Agent} terminates the investigation once all hypotheses have been examined and directs (EC)2 to the investigation conclusion stage.

\vspace{-4mm}
\subsubsection{Investigation Conclusion:} \label{subsubsec:approach:conclude_investigation}
The \textit{Explanation Agent} is responsible for constructing a comprehensive natural language report that documents the entire investigation process.
The agent receives the full conversation, including all hypotheses investigated, the evidence gathered, and the \textit{Hypothesis Evaluator Agent's} judgments, and produces a narrative explanation that summarizes the significant investigative steps performed throughout the workflow.
For each hypothesis examined during the investigation cycle, the explanation includes: the hypothesis itself, the factual evidence retrieved in support of or against it, and the \textit{Hypothesis Evaluator Agent's} judgment.
The narrative, therefore, provides a chronological and evidence-grounded account of (EC)2's reasoning regarding the event: the investigative directions explored, the evidence uncovered, and the way in which each piece of information contributed to the final conclusion.

It should be noted that (EC)2 does not attempt to choose between conflicting hypotheses.
If two or more explanations remain viable based on the available evidence, the final report includes all of them.
The investigation framework presents the complete investigative record - supported, refuted, and unresolved hypotheses alike - leaving the final judgment to the human analyst. (EC)2 is designed to provide transparent reasoning rather than enforce a conclusion.

The \textit{Explanation Reviewer Agent} evaluates the quality and reliability of the generated report before finalization.
The reviewer receives the narrative produced by the \textit{Explanation Agent} and ensures that the explanation is complete, coherent, and faithfully reflects the evidence gathered during the investigation.
This includes checking that all hypotheses, even those that remain unresolved or conflict with one another, are accurately represented.
The \textit{Explanation Reviewer Agent} also verifies that the narrative contains no speculative claims and that the structure is clear and logically consistent. If gaps or inconsistencies are identified, the agent requests revisions, prompting the \textit{Explanation Agent} to refine the report.
This refinement loop continues until the reviewer determines that the explanation is correct and sufficiently detailed to serve as the final output.

\vspace{-4mm}
\section{Experimental Setup} 
\label{sec:experimentalSetup}
% \vspace{-2mm}
\subsection{Dataset}
The main evaluation dataset used in this work is the CSE-CIC-IDS2018 dataset \cite{sharafaldin2018toward}, a collaborative effort between the Communications Security Establishment (CSE) and the Canadian Institute for Cybersecurity (CIC).
The dataset was generated in a controlled AWS-hosted network environment comprising 50 attacking machines and a victim organization with 420 machines and 30 servers across five departmental subnets.
It was designed to address the limitations of prior IDS benchmarks by ensuring complete traffic capture, diverse attack coverage, and realistic benign background traffic generated through statistical user behavior profiling.
The dataset covers seven attack categories: brute-force, DoS, DDoS, web, botnet, Heartbleed and infiltration attacks.

Data was recorded over 10 days, with benign traffic captured throughout and attacks executed on designated days according to a predefined schedule.
For each recorded day, the dataset contains raw network traffic in PCAP format, per-machine system event logs, and labeled flow records extracted from the raw captures using CICFlowMeter-V3.
% The flow records are bidirectional, with over 80 statistical features covering packet lengths, inter-arrival times, flag counts, flow duration, and related traffic characteristics, and are organized as per-day CSV files.
Each flow is labeled according to its attack category, determined by matching its timestamp and IP addresses against the documented attack schedule, which specifies the time window and participants of each attack.
The raw PCAP files contain approximately 1.2 billion packets, providing full packet-level visibility into the captured traffic. The dataset contains a total of approximately 16 million flow records, of which roughly 17\% are labeled as attack traffic and the remainder as benign.

\vspace{-4mm}
\subsection{Implementation Details}
(EC)2 requires two data sources to support its investigation pipeline: a network topology knowledge graph and a traffic database.
In this section, we describe how both were constructed from the CSE-CIC-IDS2018 dataset, which provides the raw network traffic captures and host information needed to build them.

The network topology knowledge graph is constructed in two steps.
First, the physical topology is derived by combining information from two sources.
From the raw PCAP files, device-level attributes are extracted through packet inspection, including IP addresses, MAC addresses, and hostnames.
From the dataset documentation, additional context is obtained: device types, attacker IP identification, internal-to-public IP address mappings, subnet-to-department assignments, and the overall network hierarchy.
Routers, switches, and firewalls are added manually based on the documented network hierarchy to connect devices into a hierarchical structure reflecting the organization's network, and the result is serialized as a JSON file.
In the second step, the JSON file is ingested into a Neo4j graph database, where devices, IP addresses, MAC addresses, and departments are represented as typed nodes connected by relationships that capture device connectivity, addressing, and organizational affiliation. This structure supports efficient querying by the \textit{Knowledge Graph Agent}.

The network traffic database consists of two complementary representations of the network data, both stored in a PostgreSQL database.
The first is a packet-level index derived from the raw PCAP files using tshark~\cite{tshark}, where each row corresponds to a single packet and captures only structured, searchable header fields.
This design is motivated by storage constraints, as storing full packet content at this scale would result in an extremely large table, and by query efficiency, as the reduced schema lowers search time across the table.
Each indexed packet maintains a reference to its source via a (pcap\_id, frame\_number) pair, allowing full packet reconstruction directly from the original PCAP files when deeper inspection is required.
%This design enables efficient bulk querying over large packet sets while preserving access to complete packet-level evidence.
The second representation is a flow-level table constructed from the NetFlow CSV records provided with the dataset, where each row represents an aggregated bidirectional flow summarizing statistical and behavioral properties such as packet counts, byte rates, inter-arrival times, and TCP flag counts.
Both tables are indexed on commonly queried fields such as IP addresses, ports, and timestamps to support efficient retrieval.

The \textit{Traffic Agent} has access to both representations and selects the appropriate level of detail based on the investigative need. Flow-level records are preferred by default, as they capture aggregated traffic behavior at lower computational cost. Packet-level data is reserved for cases requiring fine-grained inspection of individual network events. In practice, this design was reflected in our experiments, where the agent predominantly queried the flow-level table, with packet-level access occurring selectively for events requiring deeper inspection.
All agents in the (EC)2 framework are powered by GPT-5-mini.

\vspace{-4mm}
\section{Evaluation} \label{sec:evaluation}
\begin{figure*}[htp]
\centering
\includegraphics[width=\textwidth]{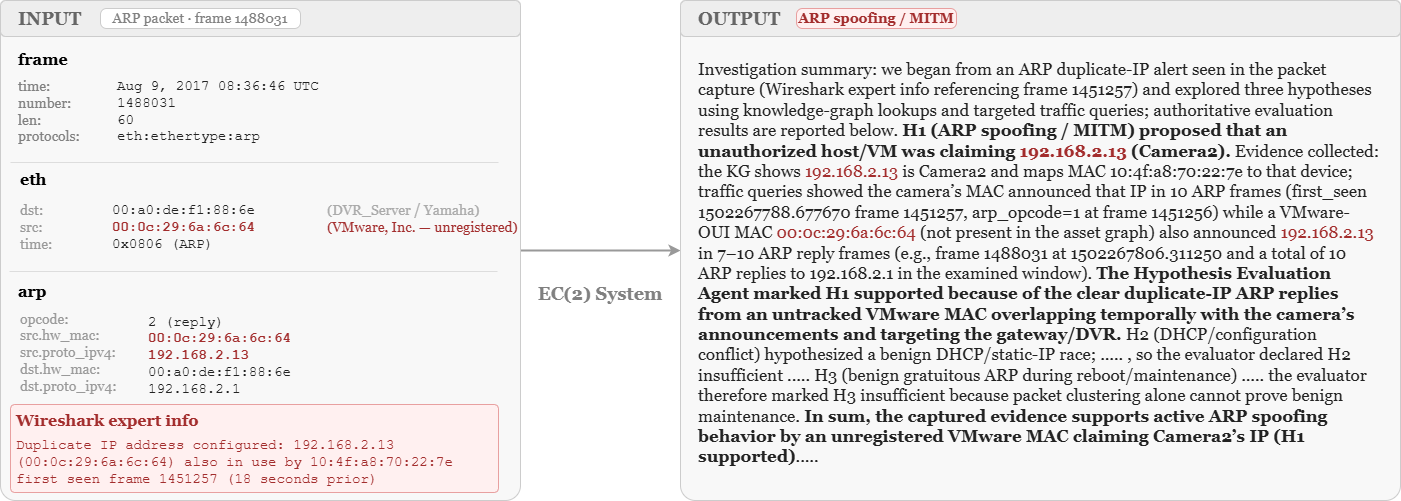} % full page width
\caption{Example of the (EC)2 framework's input and output. Left: a raw ARP packet. Right: the natural language report of the investigation summary.}
\label{fig:ec2_input_output}
\end{figure*}

% \vspace{-2mm}
\Cref{fig:ec2_input_output} presents an example of the framework's input and output: a raw ARP packet flagged for a duplicate-IP conflict serves as input, and the output is a corresponding natural language report of the investigation summary identifying the event as a high-confidence ARP spoofing and targeted cache-poisoning attempt.

In this research, our evaluation aims to answer three research questions:
\begin{itemize}
\item \textbf{RQ1:} How does (EC)2 compare to a state-of-the-art LLM-based explainability framework for network intrusion detection in terms of explanation quality across multiple attack scenarios (comparison with baseline method)?
\item \textbf{RQ2:} To what extent do (EC)2's individual components contribute to the quality of its explanations (ablation study)?
\item \textbf{RQ3} Can (EC)2 improve network intrusion detection performance (detection enhancement)?
\end{itemize}

All LLM-based judges used throughout the evaluation are implemented using GPT-5.1.

\subsection{RQ1: Comparison with Baseline Method} \label{subsec:evaluation:rq1}
eX-NIDS~\cite{houssel2026ex} is a hybrid LLM-based framework designed to enhance the interpretability of flow-based \acp{NIDS}.
% Its Prompt Augmenter module receives the NetFlow record as input and enriches it with three categories of contextual information: NetFlow feature specifications, IP-specific knowledge including geolocation and threat intelligence, and protocol identification mappings.
% This augmented prompt is submitted to an LLM in a single inference call to produce a natural language explanation of why the flagged flow is considered malicious.
Its Prompt Augmenter module enriches NetFlow records with contextual information - NetFlow feature specifications, IP-specific knowledge, and protocol identification mappings before submitting the augmented prompt to an LLM in a single inference call to produce a natural language explanation of why the flagged flow is considered malicious.
The authors publicly released their generated explanations on GitHub~\cite{exnidsGithub}, allowing us to directly compare our framework's explanations with their published explanations for 50 anomalous NetFlow events from the CSE-CIC-IDS2018 dataset~\cite{sharafaldin2018toward}.

\begin{figure*}[htp]
\centering
\includegraphics[width=\textwidth]{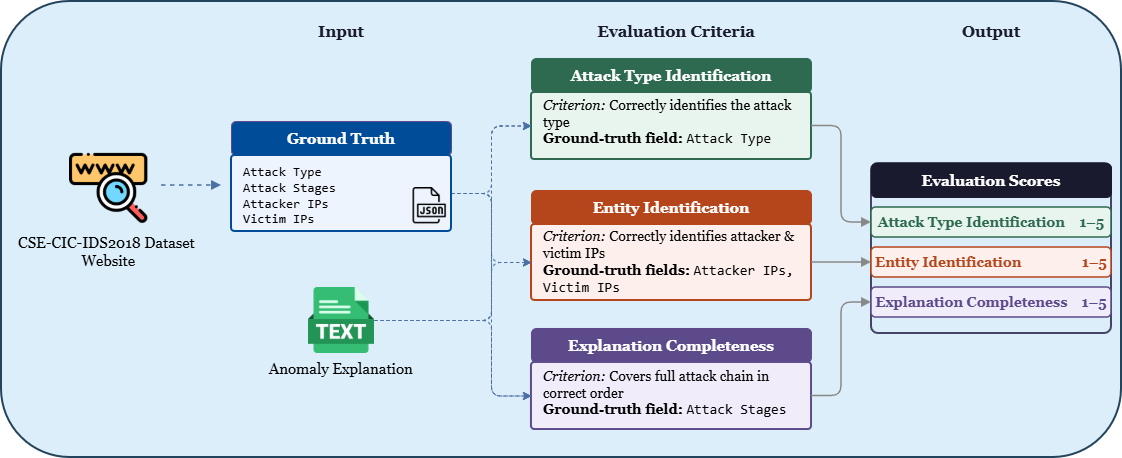} 
\caption{Explanation quality assessment methodology.}
\label{fig:evaluation_diagram}
\end{figure*}

To address RQ1, we compare (EC)2's explanations to eX-NIDS on these 50 events. 
%, ensuring a controlled and fair comparison across diverse attack types.
We evaluate explanation quality using three criteria, each scored on a 1-5 scale by an LLM-based evaluator configured as a senior security analyst, as illustrated in \Cref{fig:evaluation_diagram}.
% As can be seen in \Cref{fig:evaluation_diagram}, which presents the explanation quality assessment methodology, the evaluator receives both the generated explanation and a structured ground truth, and scores each criterion independently.

% % For each anomalous event, the explanation is compared to a ground truth comprising the attack type, attacker and victim IPs, and the ordered attack stages.
% % Three criteria are scored independently by an LLM-based evaluator and manually reviewed for reliability.
% For each event, the explanation is compared to a ground truth constructed from the publicly available dataset documentation, which includes descriptions of each attack scenario and the roles of the participating entities. The ground truth comprises the attack type, attacker and victim IPs, and the ordered attack stages.
% Three criteria are scored independently and manually reviewed for reliability:
For each explanation, the evaluator scores each criterion independently, using the relevant portion of the ground truth, which comprises the attack type, attacker and victim IPs, and the ordered attack stages - constructed from the publicly available dataset documentation.
Three criteria are scored independently and manually reviewed for reliability:
\textbf{Attack Type Identification} assesses whether the explanation correctly identifies the specific category of attack that occurred.
The ground-truth field is the attack type label.
\textbf{Entity Identification} assesses whether the explanation correctly identifies the attacker and victim IP addresses and characterizes their respective roles in the event.
The ground-truth fields are the attacker and victim IP addresses.
\textbf{Explanation Completeness} assesses whether the explanation captures the full attack chain, including all key stages and their correct temporal sequence.
The ground-truth field is the ordered sequence of attack stages.

Scores range from one (incorrect or entirely missing) to five (fully correct, complete, and accurate), with intermediate values reflecting partial correctness.
To mitigate stochastic variability, each explanation was scored across three independent evaluation runs and the reported scores are cumulative means.
The eX-NIDS scores were obtained by applying our LLM-based evaluator to the explanations of Houssel et al. for the same 50 events, ensuring that both systems were assessed under identical evaluation conditions.

\vspace{-4mm}
\begin{figure}[htp]
    \centering
    \begin{minipage}{0.48\linewidth}
    \centering
    \includegraphics[width=\linewidth]{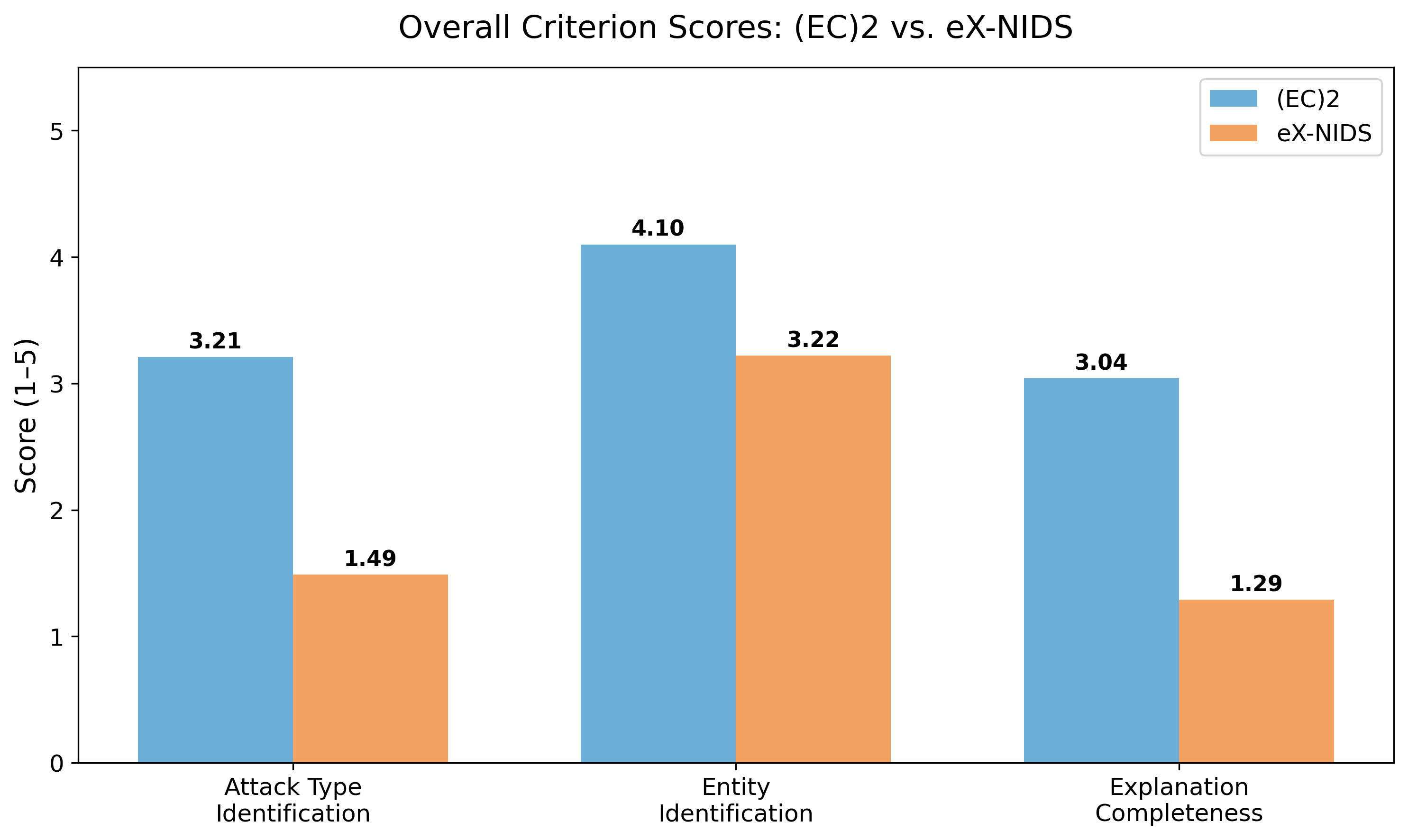}
    \caption{Overall mean explanation quality scores for (EC)2 and eX-NIDS per evaluation criterion.}
    \label{fig:bar_plot}
    \end{minipage}
    \hfill
    \begin{minipage}{0.48\linewidth}
    \centering
    \includegraphics[width=\linewidth]{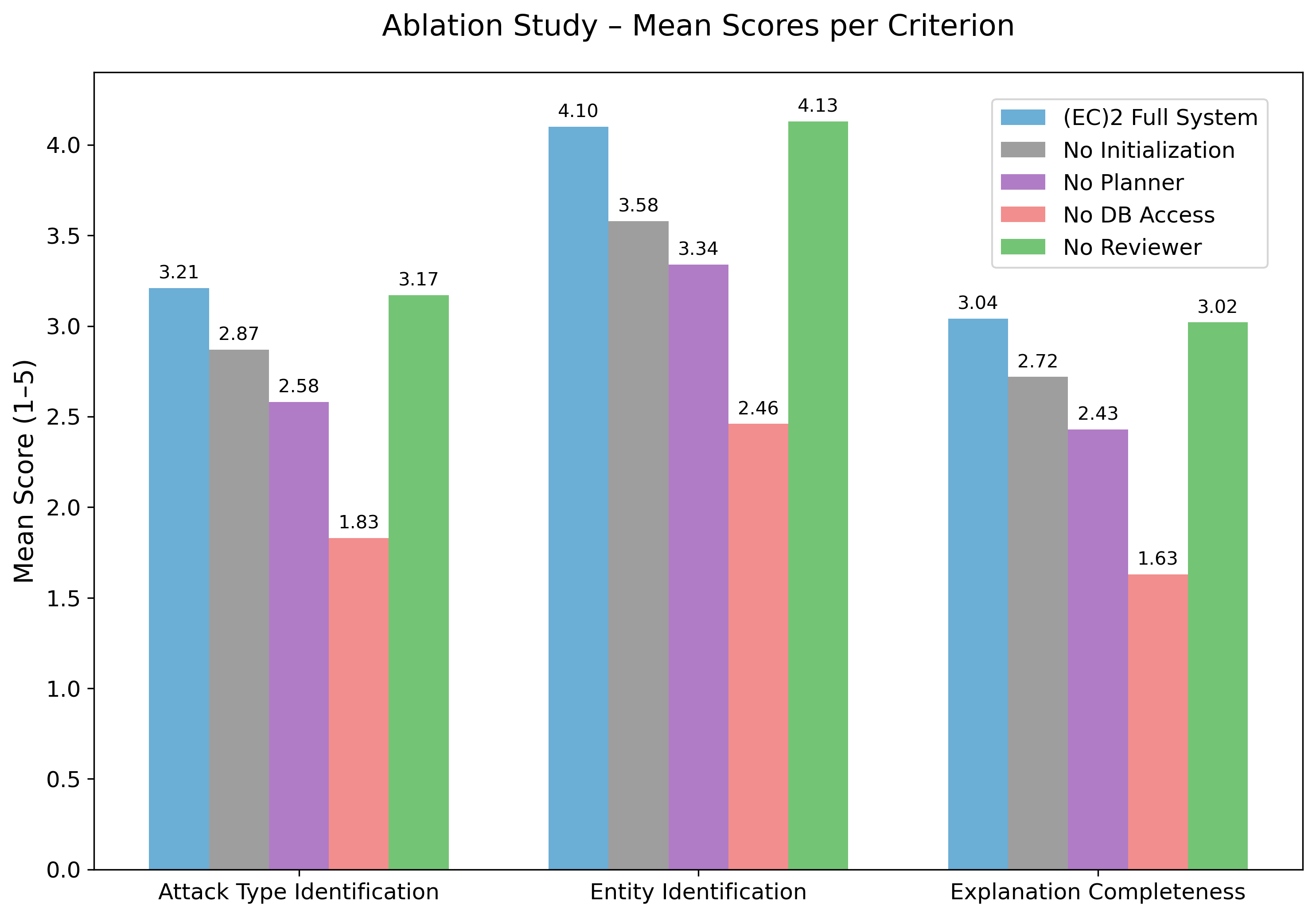}
    \caption{Mean evaluation scores per criterion for each (EC)2 ablation variants.}
    \label{fig:ablation_graph}
    \end{minipage}
\end{figure}

\vspace{-4mm}
\Cref{fig:bar_plot} presents the overall mean scores across all 50 events.
(EC)2 outperforms eX-NIDS on all three criteria.
% The most pronounced advantage is in Attack Type Identification, where (EC)2's score of 3.21 is more than double that of eX-NIDS, which achieved a score of 1.49.
% This reflects the structural advantage of hypothesis-driven, evidence-grounded investigation: (EC)2 actively queries network topology and traffic databases to contextualize the event and characterize the attack type as part of a structured reasoning process, whereas eX-NIDS generates an explanation in a single inference step from a pre-enriched prompt.
The most pronounced advantage is in Attack Type Identification ((EC)2 3.21 vs. eX-NIDS 1.49), reflecting the structural advantage of hypothesis-driven, evidence-grounded investigation: (EC)2 actively queries network topology and traffic databases to contextualize and characterize the attack type, whereas eX-NIDS generates an explanation in a single inference step.
Without iterative evidence retrieval and validation, eX-NIDS frequently fails to identify the correct attack category, particularly for attack types requiring cross-event reasoning or multi-stage context.
% The narrowest gap (4.10 vs. 3.22) between the methods is seen on the Entity Identification criterion; this is expected, as both systems have access to IP-level information - eX-NIDS through threat intelligence and geolocation lookups and (EC)2 through its \textit{Knowledge Graph Agent}.
% However, (EC)2 more reliably identifies and articulates entity roles by explicitly reasoning about device relationships within the network topology rather than relying solely on external reputation lookups.
Entity Identification shows the narrowest gap (4.10 vs. 3.22), as both systems have access to IP-level information, though (EC)2 more reliably identifies entity roles by reasoning about device relationships within the network topology.
% The largest relative improvement is in Explanation Completeness (3.04 vs. 1.29), the most demanding criterion, requiring reconstruction of the full attack chain in the correct sequence.
% eX-NIDS, operating from a single enriched prompt, often characterizes traffic at a high level but fails to reconstruct multi-stage attack progression.
% (EC)2's iterative hypothesis-testing cycle, which continues until all plausible investigative directions have been explored, enables it to capture a substantially more complete picture of how the attack unfolded.
The largest relative improvement is in Explanation Completeness (3.04 vs. 1.29): eX-NIDS often characterizes traffic at a high level but fails to reconstruct multi-stage attack progression, whereas (EC)2's iterative hypothesis-testing cycle enables it to capture a substantially more complete picture of how the attack unfolded.

% \Cref{fig:convergence} presents test-retest results confirming score stability across three independent evaluation runs.
% (EC)2's scores converge rapidly and remain stable across runs for all three criteria.
% eX-NIDS scores are similarly stable at their lower baseline.
% The consistency of the gap across runs indicates that the performance difference is not an artifact of evaluation variance but reflects a structural property of each system's explanation quality.

\begin{figure*}[htp]
\centering
\includegraphics[width=\textwidth]{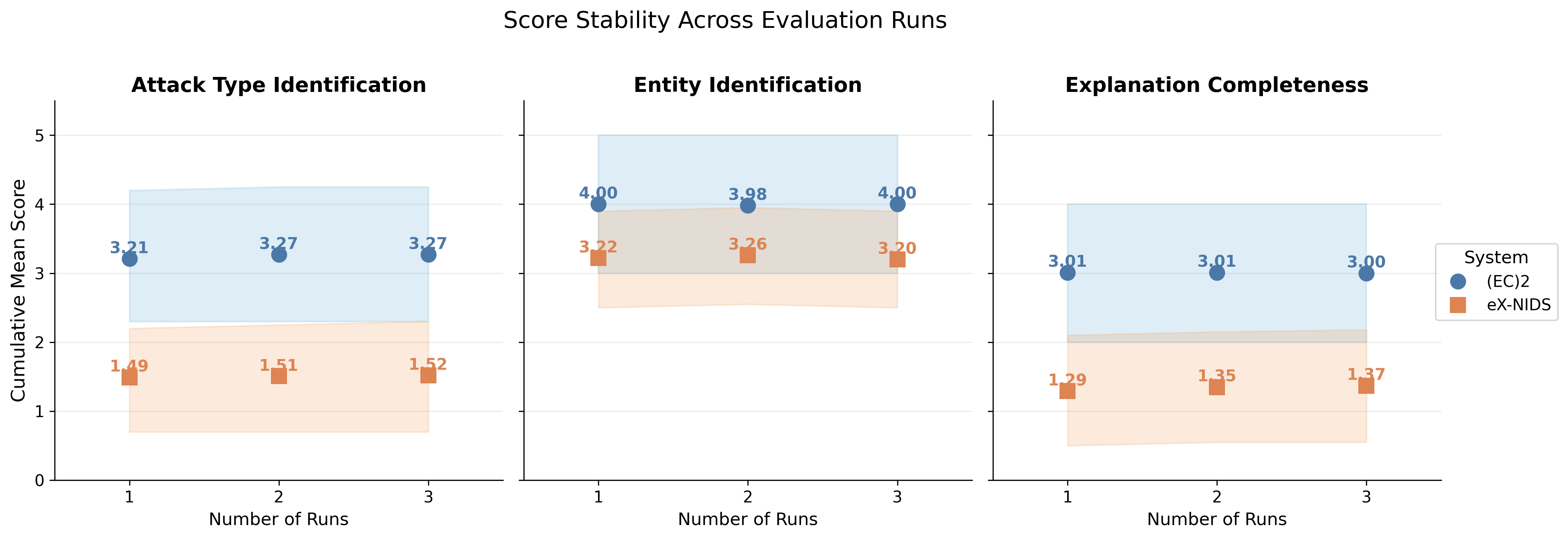} % full page width
\caption{Score stability across three independent evaluation runs for (EC)2 and eX-NIDS.}
\label{fig:convergence}
\end{figure*}

% Score stability across three independent evaluation runs was confirmed for both systems (see \Cref{fig:convergence} in the appendix), demonstrating that the evaluation scores are consistent and reproducible, and that the observed performance gap between (EC)2 and eX-NIDS is attributable to differences in explanation quality rather than stochastic variability in the LLM-based evaluation process.
Score stability across three independent evaluation runs was confirmed for both systems (\Cref{fig:convergence}), demonstrating that the observed performance gap is attributable to differences in explanation quality rather than stochastic variability in the LLM-based evaluation process.

% Together, these results directly address RQ1: (EC)2 produces explanations of measurably higher quality than eX-NIDS across all evaluation criteria and attack categories.
% The comparison demonstrates that moving from prompt-augmented, single-call LLM explanation toward autonomous, hypothesis-driven, multi-agent investigation represents a qualitatively different level of explainability, one that more closely mirrors the depth and structure of real analyst investigation workflows.
Together, these results directly address RQ1: (EC)2 produces explanations of measurably higher quality than eX-NIDS across all evaluation criteria and attack categories, demonstrating that autonomous, hypothesis-driven, multi-agent investigation produces substantially better explanations than prompt-augmented, single-call LLM explanation.

\vspace{-4mm}
\subsection{RQ2: Ablation Study} \label{subsec:evaluation:rq2}
% To assess the contribution of individual components to (EC)2's explanatory quality, we conducted an ablation study structured around two complementary sub-questions.
% Both sub-studies were applied to the same 50 events used in RQ2, with explanations evaluated using the same three criteria: Attack Type Identification, Entity Identification, and Explanation Completeness, scored on a 1-5 scale by an LLM-based evaluator across three independent runs.
To assess the contribution of individual components to the quality of (EC)2's explanations, we conducted an ablation study with two complementary subquestions, evaluated on the same 50 events and three criteria as RQ1.

\textbf{RQ2.1: Which agents in the (EC)2 pipeline are necessary for producing high-quality explanations?}
% To address this, four variants of the system were evaluated, each with a distinct agent removed:
% \begin{itemize}
%     \item \textbf{No Initialization:} The \textit{Contextual Questions Generator Agent} and its evidence retrieval step are removed. The \textit{Hypothesis Generator Agent} receives only the raw anomalous event, with no prior factual context.
%     \item \textbf{No Planner:} The \textit{Planner Agent} is removed. The \textit{Evidence Gathering Agent} receives each hypothesis directly and is responsible for both determining which questions are necessary and retrieving answers, without a dedicated planning step.
%     \item \textbf{No Database Access:} The \textit{Evidence Gathering Agent} relies solely on the LLM's internal knowledge, without querying any external data source.
%     \item \textbf{No Reviewer:} The \textit{Explanation Reviewer Agent} is removed, and the narrative produced by the \textit{Explanation Agent} is finalized without any quality validation or revision cycle.
% \end{itemize}
Four variants of (EC)2 were evaluated, each with a distinct component removed: \textbf{No Initialization} (\textit{Contextual Question Generator Agent} removed), \textbf{No Planner} (\textit{Planner Agent} removed), \textbf{No Database Access} (\textit{Evidence Gathering Agent} relies solely on LLM internal knowledge), and \textbf{No Reviewer} (\textit{Explanation Reviewer Agent} removed).

% \textbf{No Initialization:} The \textit{Contextual Questions Generator Agent} and its evidence retrieval step are removed. The \textit{Hypothesis Generator Agent} receives only the raw anomalous event, with no prior factual context.
% \textbf{No Planner:} The \textit{Planner Agent} is removed. The \textit{Evidence Gathering Agent} receives each hypothesis directly and is responsible for both determining which questions are necessary and retrieving answers, without a dedicated planning step.
% \textbf{No Database Access:} The \textit{Evidence Gathering Agent} relies solely on the LLM's internal knowledge, without querying any external data source.
% \textbf{No Reviewer:} The \textit{Explanation Reviewer Agent} is removed, and the narrative produced by the \textit{Explanation Agent} is finalized without any quality validation or revision cycle.

%\vspace{-4mm}
\begin{table}[htp]
    \centering
    \small
    \caption{Ablation study results: all variants.}
    \label{tab:ablation_results}
    \resizebox{\linewidth}{!}{%
    \begin{tabular}{|l|c|c|c|c|c|c|c|c|c|}
    \hline
    \rowcolor{cyan!10}
    \textbf{Framework Variant} &
    \multicolumn{2}{c|}{\textbf{Attack Type Identification}} &
    \multicolumn{2}{c|}{\textbf{Entity Identification}} &
    \multicolumn{2}{c|}{\textbf{Explanation Completeness}} \\
    \hline
     & Mean & p-value
     & Mean & p-value
     & Mean & p-value \\
    \hline
    (EC)$^2$ Full System
    & 3.21 & --
    & 4.10 & --
    & 3.04 & -- \\
    \hline
    No Reviewer
    & 3.17 & 0.530
    & 4.13 & 0.634
    & 3.02 & 0.746 \\
    \hline
    No Planner
    & 2.58 & $< 0.001$
    & 3.34 & $< 0.001$
    & 2.43 & $< 0.001$ \\
    \hline
    No Initialization
    & 2.87 & $< 0.001$
    & 3.58 & $< 0.001$
    & 2.72 & $< 0.001$ \\
    \hline
    No DB Access
    & 1.83 & $< 0.001$
    & 2.46 & $< 0.001$
    & 1.63 & $< 0.001$ \\
    \hline
    Traffic Only
    & 2.89 & $< 0.001$
    & 3.95 & 0.017
    & 2.76 & $< 0.001$ \\
    \hline
    Topology Only
    & 2.31 & $< 0.001$
    & 2.65 & $< 0.001$
    & 2.18 & $< 0.001$ \\
    \hline
    \end{tabular}%
    }
\end{table}

We use paired t-tests to assess whether the differences between the varians and the full framework are statistically significant; the null hypothesis ($H_0$) states that removing a component produces no meaningful change in the mean scores.
$H_0$ is rejected for No Database Access, No Planner, and No Initialization across all criteria (p < 0.001), but it cannot be rejected for No Reviewer (p > 0.05).
\Cref{tab:ablation_results} (rows 2-5) presents the mean scores and p-values for each variant; \Cref{fig:ablation_graph} visualizes the score distributions for the variants per each criterion.
The most severe degradation is observed for the \textbf{No Database Access} variant: without external data, the framework cannot anchor reasoning in observed network behavior, rendering explanations unreliable regardless of the hypothesis or planning quality.
The \textbf{No Planner} and \textbf{No Initialization} variants each result in significant declines, confirming that dedicated planning and upfront context gathering are both necessary for targeted evidence retrieval.
The No Initialization variant degrades less severely than the No Planner variant, as the two-round hypothesis-generation structure enables partial recovery through evidence gathered in the first round.
The \textbf{No Reviewer} variant shows no meaningful difference, indicating that the \textit{Explanation Agent}'s output is sufficiently coherent that the \textit{Explanation Reviewer Agent}'s refinement loop yields no measurable improvement.

% Taken together, the RQ2.1 results confirm that the Planner Agent, the Contextual Initialization step, and database access each make a statistically significant and meaningful contribution to (EC)2's explanation quality. The \textit{Explanation Reviewer Agent}, while part of the pipeline, does not produce a statistically detectable effect on the evaluated criteria.

% Overall, these results confirm that the \textit{Planner Agent}, Contextual Initialization, and database access each make a statistically significant contribution to explanation quality. The \textit{Explanation Reviewer Agent} does not produce a detectable effect.

\textbf{RQ2.2: How Does the Availability of Individual Evidence Sources Affect the Quality of (EC)2's Explanations?}
% RQ2.1 established that database access is critical to explanation quality.
% To address RQ2.2, two additional variants were evaluated in which only one source was made available to the system: \textbf{Topology Only}, in which the \textit{Evidence Gathering Agent} has access solely to the network topology graph, and \textbf{Traffic Only}, in which it has access solely to the network traffic database.
% These variants are evaluated alongside the No DB Access condition from RQ2.1 and the full (EC)2 framework.
% All variants are compared against the full system using paired t-tests under the same null hypothesis H\textsubscript{0} as in RQ2.1.
Building on the above mentioned finding that database access is critical, to answer RQ2.2, we evaluate two additional variants: \textbf{Topology Only} (Topology Knowledge Graph access only) and \textbf{Traffic Only} (Traffic Database access only), which their performance is compared to the No Database Access variant and the full framework, using the same paired t-test procedure.

\Cref{tab:ablation_results} (last two rows) presents the results for each variant for the three evaluation criteria.
For \textbf{Topology Only}, applying the same $H_0$ as above, the null hypothesis is rejected across all criteria ($p < 0.001$). Without traffic data, the framework lacks the communication patterns and addressing information of the attack event, resulting in score degradation across all three criteria: Attack Type Identification, Entity Identification, and Explanation Completeness.
For \textbf{Traffic Only}, $H_0$ is rejected across all criteria ($p < 0.001$). Traffic data captures the core behavioral evidence, but the absence of topology data removes organizational context (e.g., device roles and legitimate address ranges) that grounds reasoning in the broader network environment.
Topology data is thus required even when traffic data is available.
Together, both data sources are necessary for optimal quality.
Traffic data is the more critical, as its removal results in greater score degradation, while topology data provides complementary context whose contribution is reflected in the lower scores observed for the Traffic Only variant.

\subsection{RQ3: Detection Enhancement}
% Anomaly detectors classify network flows by comparing an anomaly score against a learned threshold.
% While this approach can achieve strong overall performance, it is inherently least reliable near the decision boundary - the region where the model's confidence is lowest, and misclassification risk is highest.
% In practice, this boundary region is a primary source of false positives: benign flows are flagged as malicious due to marginal deviations from normal behavior.
% Rather than treating these uncertain cases as an unavoidable limitation of the detector, we propose leveraging (EC)2's explainability-driven reasoning to reclassify them.
% By routing low-confidence cases through a full multi-agent investigation, we examine whether the depth of (EC)2's contextual analysis can recover correct classifications that the base model alone could not reliably produce, thereby improving the overall detection performance of the anomaly detection pipeline.

Anomaly detectors classify by comparing an anomaly score against a learned threshold.
This approach is inherently least reliable near the decision boundary (where model confidence is lowest and misclassification risk is highest), and in practice this boundary region is a primary source of false positives.
% Rather than treating these uncertain cases as an unavoidable limitation, we propose leveraging (EC)2's explainability-driven reasoning to reclassify them, examining whether its contextual analysis can produce correct classifications that the base anomaly detector alone could not reliably produce.
Rather than treating these uncertain cases as an unavoidable limitation, we propose leveraging (EC)2's reasoning to reclassify them, examining whether its contextual analysis can recover correct classifications that the base model alone could not reliably produce.

% To address RQ3, we apply the AE-IDS ~\cite{li2020building}, an unsupervised intrusion detection system designed for network anomaly detection.
% We run AE-IDS using the authors' published code on the CSE-CIC-IDS 2018 dataset; unlike the original evaluation, which reports results per attack scenario, we treat the full dataset as a single unified test set to reflect realistic heterogeneous traffic conditions.
% AE-IDS computes an anomaly score per NetFlow record as the \ac{RMSE} across feature-group autoencoders, with the detection threshold set to the maximum RMSE observed over normal samples during training.
% From the full test set, we identify the uncertain zone as the 5\% of NetFlow records with the smallest margin from this threshold.
% Due to the computational cost of (EC)2's multi-agent investigation, we sample 1,000 NetFlow records from this zone in a stratified manner, balanced across malicious and benign ground-truth labels.
% Each sampled record is passed to (EC)2 for investigation, and the resulting natural language explanation is submitted to an LLM judge prompted to issue a binary verdict: malicious or benign, based solely on the explanation content.
% This verdict replaces AE-IDS's original classification for that record.
% We evaluate classification performance before and after reclassification on this 1,000-record sample using McNemar's test to assess whether the improvement is statistically significant. %, as illustrated in \Cref{fig:detection_evaluation}.
To address RQ3, we apply AE-IDS \cite{li2020building}, an unsupervised intrusion detection system that computes an anomaly score per NetFlow record as the mean RMSE across feature-group autoencoders, with the detection threshold set to the maximum RMSE observed over normal samples during training.
We run AE-IDS on the CSE-CIC-IDS2018 dataset, treating the full dataset as a single unified test set to reflect realistic heterogeneous traffic conditions.
From the full test set, we identify the uncertain zone as the 5\% of NetFlow records with the smallest margin from the threshold.
Due to the computational cost of (EC)2's multi-agent investigation, we sample 1,000 NetFlow records from this zone in a stratified manner, balanced across malicious and benign ground truth labels.
Each sampled record is passed to (EC)2 for investigation, and the resulting explanation is submitted to an LLM prompted to issue a binary verdict: malicious or benign, based solely on the explanation content, replacing AE-IDS's original classification for that record.
% (see Appendix for evaluation pipeline illustration).

\begin{comment}
\begin{figure*}
\centering
\includegraphics[width=\textwidth]{Images/classification_evaluation_pipeline.png} % full page width
\caption{Evaluation pipeline for detection enhancement using (EC)2 explanations.}
\label{fig:detection_evaluation}
\end{figure*}
\end{comment}

Li et al. reported per-scenario recall values for AE-IDS; we compute a weighted average
% of these per-scenario values, 
yielding an overall recall of \textbf{0.63}. Re-running AE-IDS on the full unified dataset yields a recall of \textbf{0.58}, a moderate drop, as expected when a single shared threshold must simultaneously generalize across heterogeneous attack types.
Within this unified test set, the uncertain zone 
% (5\% of records with the smallest margin from the threshold)
yields a stratified sample of 1,000 records (500 malicious, 500 benign).
% On this sample, AE-IDS achieves Recall of \textbf{0.54}, Precision of \textbf{0.56}, and FPR of \textbf{0.46}, a substantial degradation relative to its overall performance, confirming that boundary-region cases are significantly harder to classify correctly.
% After reclassification using (EC)2 explanations, Recall improves to \textbf{0.72}, Precision to \textbf{0.75}, and FPR drops to \textbf{0.27}.
% To assess statistical significance, we apply McNemar's test, which focuses on the cases where the two classifiers disagree - one gets it right while the other gets it wrong - and asks whether one is systematically fixing more mistakes than the other.
% The null hypothesis is that there is no systematic advantage of one classifier over the other.
% Of the 1,000 cases, (EC)2 corrected \textbf{207} misclassifications made by AE-IDS while introducing errors in only \textbf{37} previously correct cases.
% The resulting test statistic ($\chi^2 = 111.8,\; p < 0.001$) leads us to reject the null hypothesis, confirming that the improvement is statistically significant and cannot be attributed to chance.
On this sample, AE-IDS achieves a recall of \textbf{0.54}, a precision of \textbf{0.56}, and a \ac{FPR} of \textbf{0.43}, confirming that boundary-region cases are significantly harder to classify correctly.

After reclassification using (EC)2 explanations, recall improves to \textbf{0.72}, precision to \textbf{0.73}, and FPR drops to \textbf{0.26}. We apply McNemar's test, which focuses on cases where two classifiers disagree and asks whether one is systematically fixing more mistakes than the other.
$H_0$ states that neither classifier has a systematic advantage.
Of the 1,000 sampled records, AE-IDS correctly classified \textbf{558} and misclassified \textbf{442}. (EC)2 corrected \textbf{207} of the misclassified cases while introducing errors in only \textbf{37} previously correct cases, yielding $\chi^2 = 118.4,\; p < 0.001$.
This leads us to reject $H_0$, confirming that the improvement is statistically significant and cannot be attributed to chance.

% These results provide a clear answer to RQ3. (EC)2 improves network intrusion detection performance beyond what AE-IDS achieves alone, specifically in the region where anomaly detection is least reliable.
% By routing uncertain cases through a full multi-agent investigation, (EC)2 successfully reclassifies cases that the base model could not resolve, reducing both missed attacks and false alarms.
% McNemar's test confirms that this improvement is statistically significant and not due to chance.
% These findings suggest that (EC)² can serve not only as an explanation layer on top of an existing detection pipeline, but as an active component that strengthens its overall detection capability.
% While this evaluation is conducted on a sample of 1,000 records due to the computational cost of (EC)2's multi-agent investigation, the uncertain zone spans a much larger portion of real-world traffic, suggesting that the cumulative impact on full-scale deployment could be substantial.

These results directly address RQ3: (EC)2 improves detection performance beyond that of AE-IDS in the region where anomaly detection is least reliable, successfully reclassifying uncertain cases and reducing both missed attacks and false alarms,
% These findings suggest that (EC)2 can serve not only as an explanation layer but as an active layer that strengthens overall detection capability.
% While this evaluation was conducted on a 1,000-record sample due to computational cost, the uncertain zone spans a much larger portion of real-world traffic, suggesting that the cumulative impact on full-scale deployment could be substantial.
suggesting that (EC)2 can serve not only as an explanation layer but as an active component that strengthens overall detection capability.

% \subsection{Runtime Analysis}
% We also examined the end-to-end runtime of the proposed investigation system under the Full Information configuration.
% On average, generating a complete explanation required approximately 17 minutes per event.
% Although this duration is relatively long, it reflects the multi-stage investigative process implemented by (EC)2.
% As described earlier in the paper, the workflow involves generating hypotheses, retrieving external evidence, and validating investigative paths before composing a final explanation.
% The 17-minute runtime, therefore, corresponds to performing the full analytical procedure automatically, rather than executing a single model call or simple post-hoc interpretation.

% From an operational perspective, this runtime does not impose fundamental limitations on (EC)2’s usability, as investigations for different anomalous events can be executed in parallel, allowing (EC)2 to scale horizontally across available compute resources.
% In practical deployments, increasing server capacity, expanding token limits, and optimizing backend infrastructure can further reduce latency and support higher investigation throughput.
% Our internal observations suggest that the retrieval stages, specifically external data interactions and the LLM's handling of large data volumes, contribute most to the overall latency.
% Taken together, the runtime results show that (EC)2 can perform comprehensive event investigations in a fully automated manner, with a runtime that reflects the scope of the investigative process.

On average, generating a complete explanation required approximately 17 minutes per event, reflecting the multi-stage investigative process: hypothesis generation, iterative evidence retrieval, and explanation validation. This runtime does not impose fundamental operational limitations, as investigations for different events can be executed in parallel. Our observations indicate that retrieval stages - external data interactions and LLM handling of large data volumes - contribute most to overall latency.

%\vspace{-4mm}
\section{Related Work} \label{sec:relatedWork}
% \vspace{-2mm}
\subsection{LLM-Based Approaches for Cybersecurity Operations}
LLMs' integration in cybersecurity operations has received significant research attention across several domains, including network anomaly detection~\cite{ali2023huntgpt,baral2024adaptive}, phishing detection~\cite{lim2025explicate}, and DDoS attacks~\cite{chatzimiltis2025interpretable}.
These works provide the LLM with model outputs and contextual metadata, enabling it to generate more human-understandable narratives.
Some studies have gone further by incorporating deep learning interpretability tools or by using the LLM as both a classifier and an explainer \cite{balasubramanian2025anomalyexplainer,houssel2408towards,palma2025leveraging}, yet they still lack the ability to verify generated claims.
eX-NIDS~\cite{houssel2026ex} and ChatIDS~\cite{juttner2024chatids} enrich the input context before a single LLM inference call to explain IDS alerts.
In all of these methods, the LLM's primary role remains interpretive rather than investigative: they translate outputs, summarize feature importance, or provide general advisory information but do not determine the root cause or reconstruct the sequence of events underlying the event.

\begin{comment}
The most closely related work to ours in this category is eX-NIDS \cite{houssel2026ex}, proposed by Houssel et al., described in detail in \Cref{subsec:experiments:baseline_methods}.
While eX-NIDS represents a meaningful advance over raw feature importance by incorporating domain-specific context into LLM prompts, it remains a single-pass, non-iterative system: explanations are produced without hypothesis generation, without retrieval from live data sources, and without any mechanism to verify or refute proposed explanations against actual network evidence.
As we demonstrate empirically in \Cref{subsec:evaluation:rq1}, this architectural difference has measurable consequences for explanation quality, particularly on the dimensions of attack type identification and explanation completeness, where iterative multi-agent investigation yields substantially richer outputs.
\end{comment}

\subsection{Hypothesis-Driven Reasoning With LLMs}
A few studies adopted hypothesis-driven reasoning to move toward deeper causal insight. The IEM system~\cite{nikolakopoulos2024large} enhances law-enforcement investigations through a RAG-enabled LLM trained on criminology data, generating narrative crime reconstructions and potential suspects with associated confidence scores.
A related conceptual foundation exists in cybersecurity threat hunting, where analysts propose a potential threat and iteratively gather and evaluate evidence to validate or reject it, aiming to uncover correlational and causal relationships between the suspected threat and the targeted asset~\cite{wafula2019carve}.

\subsection{Multi-Agent LLM Systems for Cyber and Digital Forensics}
Multi-agent LLM architectures have also been applied to digital forensics.
Wickramasekara and Scanlon~\cite{wickramasekara2024framework} proposed a multi-agent AutoGen-based framework supporting the examination, analysis, and reporting stages of forensic practice, but the investigation advances only when a human investigator initiates questions or tasks.
PROVSEEK \cite{mukherjee2025llm} is a more automated LLM-driven multi-agent system for provenance-based forensic analysis that reconstructs complex attacks in a scalable, interpretable way, although its investigation path is shaped by external CTI reports that seed the system.
%In contrast, our framework constructs its own investigative trajectory by generating hypotheses, determining what evidence is needed, and autonomously conducting the investigation end-to-end.

\begin{comment}
Taken together, existing work demonstrates clear progress toward more interpretable security systems, yet a substantial gap remains between explaining a model’s decision and explaining the event itself.
Feature-based XAI systems translate model outputs into natural language but do not determine the underlying cause of the anomaly.
Meanwhile, emerging hypothesis-driven or agent-based investigation frameworks depend either on human-initiated queries or on external structures such as \ac{CTI} reports to guide the investigative path.
\end{comment}

To the best of our knowledge, our work is the first to introduce a fully automated, hypothesis-driven investigation system for network forensics - one that does not rely on feature explanations, user prompts, or predefined investigation templates.
Instead, our approach autonomously constructs hypotheses, determines the required evidence, queries relevant data sources, and iteratively validates or rejects each investigative direction.

% To the best of our knowledge, our work is the first to introduce a fully automated, hypothesis-driven investigation system for network forensics—one that does not rely on feature explanations, user prompts, or pre-defined investigation templates.
% Instead, our approach autonomously constructs hypotheses, determines the required evidence, queries relevant data sources, and iteratively validates or rejects each investigative direction.
%By shifting from passive interpretation to active reasoning and evidence-gathering, our system delivers end-to-end forensic explanations that uncover why an event is anomalous and how the suspicious behavior unfolded.

%\vspace{-4mm}
\section{Conclusions and Future Work} \label{sec:conclusions}
% \vspace{-2mm}
This work introduced an event-centric approach to cybersecurity explainability that shifts the focus from explaining model decisions to explaining security events themselves through hypothesis-driven, multi-agent investigation.
Our evaluation demonstrated that this approach significantly outperforms a state-of-the-art \ac{LLM}-based explainability method in explanation quality while providing explanations grounded in verifiable evidence.
(EC)2's reliance on structured data retrieval through \ac{RAG} mechanism proved essential for generating trustworthy, actionable explanations.
The investigative process of our framework is aligned with how security analysts actually conduct investigations, offering a promising path toward more interpretable and effective anomaly detection in SOC environments.

Beyond explanation quality, our results demonstrate that (EC)2 can actively improve detection performance. By reclassifying uncertain boundary-region cases using its investigation output, (EC)2 significantly reduced both missed attacks and false alarms, suggesting that the framework can serve not only as an explanation layer but as an active component that strengthens overall detection capability.

A key constraint for broad deployment remains its execution time.
The current implementation requires roughly 17 minutes per investigation.
While fully automated, this runtime makes (EC)2 more suitable for in-depth analysis of select, high-priority alerts rather than large alert streams.
Future work could focus on reducing latency by improving retrieval efficiency, caching repeated queries, exploring smaller, fine-tuned models, and developing adaptive investigation depth.
In addition, a lightweight triage layer could be integrated to identify which alerts warrant full investigation.

\begin{comment}
\section*{Code and Data Availability}
The Kitsune Network Attack Dataset~\cite{mirsky2018kitsune} used in this article was publicly released by its respective authors and can be found on Kaggle.\footnote{\url{https://www.kaggle.com/datasets/ymirsky/network-attack-dataset-kitsune}}
The code cannot be placed in the public domain; however, organizations involved in this research are open to reviewing any disclosure requests submitted through the corresponding author.
\end{comment}

%
% ---- Bibliography ----
%
% BibTeX users should specify bibliography style 'splncs04'.
% References will then be sorted and formatted in the correct style.
\newpage
\bibliographystyle{splncs04}
\bibliography{refrence}

\section*{Acronyms}
\begin{acronym}
    \acro{AI}{artificial intelligence}
    \acro{CTI}{cyber threat intelligence}
    \acro{DL}{deep learning}
    \acro{EDS}{evidence deficit score}
    \acro{IDS}{intrusion detection system}
    \acro{IPS}{intrusion prevention system}
    \acro{FPR}{false positive rate}
    \acro{LIME}{local interpretable model-agnostic explanations}
    \acro{LLM}{large language model}
    \acro{LOF}{local outlier factor}
    \acro{ML}{machine learning}
    \acro{NIDS}{network intrusion detection system}
    \acro{PDP}{partial dependence plot}
    \acro{RAG}{retrieval augmented generation}
    \acro{RMSE}{root mean square error}
    \acro{SHAP}{shapley additive explanations}
    \acro{SIEM}{security information and event management}
    \acro{SOAR}{security orchestration, automation, and response}
    \acro{SOC}{security operations center}
    \acro{TCAV}{testing with concept activation vectors}
    \acro{XAI}{explainable artificial intelligence}
\end{acronym}

%\vspace{-4mm}
%\appendix

\end{document}